\documentstyle[prl,aps,twocolumn]{revtex}

\begin{document}

\preprint{VAND-TH-98-16
\hspace{-35.0mm}\raisebox{-2.4ex}{DART-HEP-98/04}
\hspace{-4.0cm}\raisebox{-4.8ex}{IF/UERJ-DFT-13/98}
\hspace{-3.5cm}\raisebox{-7.2ex}{September 1998}}

\title{A First Principles Warm Inflation Model that Solves the
Cosmological
Horizon/Flatness Problems}

\author{Arjun Berera$^{1}$,
Marcelo Gleiser$^{2}$
and
Rudnei O. Ramos$^{3}$}

\address{
{\it $^1\;$ Department of Physics and Astronomy, Vanderbilt
University,
Nashville, TN 37235, USA}
\\
{\it $^2 \;$ Department of Physics and Astronomy, Dartmouth College,
Hanover, NH 03755, USA}
\\
{\it $^3 \;$ Univ. do Estado do Rio de Janeiro,}
{\it Inst. de F\'{\i}sica, Depto. de F\'{\i}sica Te\'orica,}
{\it 20550-013 Rio de Janeiro, RJ, Brazil}}

\maketitle

\begin{abstract}

A quantum field theory warm inflation model is presented that solves the
horizon/flatness problems. The model obtains, from the elementary
dynamics of particle physics, cosmological
scale factor trajectories that begin in a radiation dominated regime,
enter an inflationary regime and then smoothly exit back into a
radiation dominated regime, with nonnegligible radiation
throughout the evolution.

\vspace{0.34cm}
\noindent
PACS number(s): 98.80 Cq, 05.70.Ln, 11.10.Wx

\end{abstract}

\medskip

In Press Physical Review Letters 1999

\medskip

hep-ph/9809583

\bigskip

The resolution of the horizon problem, which underlies inflationary
cosmology \cite{sinf},
is that at a very early time, the
equation of state that
dictates the expansion rate of the Universe
was dominated by a vacuum energy density $\rho_v$, so
that a small causally connected patch could grow to a size that encompasses
the comoving volume which becomes the observed universe today. 

In the
standard (isentropic) inflationary scenarios, the radiation energy
density $\rho_r$ scales with the inverse
fourth power of the scale factor, becoming quickly
negligible. In such case, a short time reheating period 
terminates the inflationary period and initiates the radiation dominated
epoch. On the other hand, the only condition required by General
Relativity for inflation is that $\rho_r < \rho_v$. Inflation in the
presence of nonnegligible radiation is characterized by non-isentropic
expansion \cite{rudnei,ab2} and thermal seeds of density
perturbations \cite{bf2}. This can be realized 
in warm inflation scenarios \cite{wi} where 
there is no reheating.

The basic idea of our implementation of warm inflation is quite simple;
a scalar field, which we call the inflaton, is coupled to several other fields.
As the inflaton relaxes toward its minimum energy configuration, it will decay
into lighter fields, generating an effective viscosity. That this indeed 
happens has been demonstrated in detail in Refs. \cite{GR,hosoya,morikawa}.   
If this viscosity is large enough, the inflaton will reach
a slow-roll regime, where its dynamics becomes overdamped. This overdamped 
regime has been analyzed in Ref. \cite{BGR}. As one expects, overdamping
is most successful for the case where the inflaton is coupled to a large 
number of fields which are thermally excited, {\it i.e.}, which have
small masses compared to the ambient temperature of radiation. This result
has important consequences for cosmological applications.

In order to satisfy one of the requirements of a successful inflation (60 or so
${\rm e}$-folds), overdamping 
must be very efficient. Our goal in this Letter is to
show that it is possible to build a toy model, motivated from high energy
particle physics, that can provide enough overdamping as to produce a 
viable inflationary expansion, which smoothly exits into a radiation-dominated
regime, with $\rho_r$ slowly and monotonically decreasing throughout the
whole process. In contrast to most models in the literature,
warm inflation provides both a natural context for the slow-roll regime 
and for its graceful termination into a radiation-dominated era.

The particle physics model considered below is inspired by string theories
exhibiting $N=1$ global supersymmetry, with the inflaton coupled to
massive modes of the string, as recently discussed in Ref. \cite{ak}.
We refer the reader to this reference for details.

We thus consider the following Lagrangian of a scalar field $\phi$
interacting with $N_M\times  N_{\chi}$ scalar fields $\chi_{jk}$ and
$N_M\times  N_{\psi}$ fermion fields $\psi_{jk}$,

\begin{eqnarray}
\lefteqn{{\cal L} [ \phi, \chi_{jk}, \bar{\psi}_{jk}, \psi_{jk}] = 
\frac{1}{2}
(\partial_\mu \phi)^2 - \frac{m^2}{2}\phi^2 -
\frac{\lambda}{4 !} \phi^4} \nonumber\\
& & + \sum_{j=1}^{N_M} \sum_{k=1}^{N_{\chi}} \left\{
\frac{1}{2} (\partial_\mu \chi_{jk})^2
- \frac{f_{jk}}{4!} \chi_{jk}^4 - \frac{g_{jk}^2}{2}
\left(\phi-M_j\right)^2
\chi_{jk}^2 
\right\}
\nonumber \\
& & +\sum_{j=1}^{N_M} \sum_{k=1}^{N_{\psi}}  
\left\{ i \bar{\psi}_{jk} \not\!\partial \psi_{jk} - h_{jk} ( \phi - M_j ) 
\bar{\psi}_{jk} \psi_{jk} \right\}
\: ,
\label{Nfields}
\end{eqnarray}

\noindent 
where all coupling constants are positive: $\lambda$,
$f_{jk},g_{jk}^2, h_{jk}$ $> 0$. {}For simplicity, we consider in the following
$f_{jk}=f$, $g_{jk}=h_{jk}=g$.
Also, we will set $N_{\psi}=N_{\chi}/4$,
which along with our choice of coupling implies a cancellation
of radiatively generated vacuum energy corrections in the effective
potential \cite{dj}.  We call this kind of model a distributed mass
model (DMM), where the interaction between $\phi$ with the $\chi_{jk}$
and $\psi_{jk}$
fields establishes a mass scale distribution for the $\chi_{jk}$ and
$\psi_{jk}$ fields,
which is determined by the mass parameters $\{M\}$. Thus the $\chi_{jk}$
and $\psi_{jk}$ effective field-dependent masses, 
$m_{\chi_{jk}} (\phi,T,\{M\})$ and $m_{\psi_{jk}} (\phi,T,\{M\})$,
respectively, can be constrained
even when $\langle \phi \rangle = \varphi$ is large. A specific choice
of $\{M\}$ will be given shortly. The $\phi\chi$, $\phi\psi$ 
interactions can be
made reflection symmetric, $\phi \rightarrow -\phi$, but for our purposes
we will consider all $M_j >0$ and $\phi>0$.

The 1-loop effective equation of motion for the scalar field $\phi$ is
obtained by setting $\phi = \varphi + \eta$ in Eq. (\ref{Nfields}) and
imposing $\langle \eta \rangle=0$. Then from Weinberg's tadpole method
\cite{weiss,boya1,BGR} the 1-loop evolution equation for $\varphi$ is

\begin{eqnarray}
\lefteqn{\ddot{\varphi} + 3 H \dot{\varphi} +
m^2 \varphi + \frac{\lambda}{6} \varphi^3 +
\frac{\lambda}{2} \varphi \langle \eta^2 \rangle } \nonumber \\
& & + g^2
\sum_i^{N_M} \sum_j^{N_{\chi}} (\varphi-M_i) \langle \chi_{ij}^2 \rangle + 
g \sum_i^{N_M} \sum_j^{N_{\chi}/4} \langle \psi_{ij} \bar{\psi}_{ij} \rangle  = 0 \;.
\label{eqnew}
\end{eqnarray}

\noindent 
In the above, the term $3 H \dot{\varphi} $ describes the
energy red-shift of $\varphi$ due to the expansion of the Universe. 
In the warm-inflation regime of interest here, the
characteristic time scales (given by the inverse of the decay width) for
the fields in Eq. (\ref{Nfields}) are faster than the expansion time scale,
$1/H$. In this case, the calculation of the (renormalized) thermal
averages in Eq. (\ref{eqnew}) can be approximated just as in the Minkowski
spacetime case.  

A systematic evaluation of the averages 
in the adiabatic, strong dissipative regime was 
presented in \cite{BGR} and re-derived in \cite{YL} with
extension to fermions.  The essential
mechanism for dissipation obtained from this approach can be explained
through an intuitive kinetic theory argument first given in
\cite{hosoya} and reexamined in \cite{YL}.  (We note that the objections to 
warm inflation raised in Ref. \cite{YL} are avoided quite naturally by
coupling the inflaton to a tower of mass modes as in the present 
implementation of the DM model.) 
For the $\chi$-field averages one writes
$\langle \chi_{ij}^2 \rangle = \int (d^3 k /(2 \pi)^3)
n_{\chi_{ij}} ({\bf k}) / \omega_{\chi_{ij}} ({\bf k})$,
where the number densities $n({\bf k})$
in the strong dissipative
regime, in near equilibrium, can be written in the relaxation
time approximation to the kinetic equation as \cite{hosoya,MS} 

\begin{equation}
n ({\bf k}) \sim n^{\rm eq.} ({\bf k}) -
\tau \dot{n}^{\rm eq.} ({\bf k}).
\label{kinetic}
\end{equation}

\noindent
Here $n^{\rm eq.} ({\bf k}) = 1/(e^{\beta
\omega ({\bf k})} - 1)$ is the equilibrium number density
for $\chi$ particles,
$\omega ({\bf k}) = \sqrt{{\bf k}^2 +
m^2 (\varphi,T,\{M\})}$,
$m^2 (\varphi,T,\{M\})$ is the
effective, field dependent mass for the $\chi$-fields \cite{BGR}, and
$\tau=\Gamma^{-1}$,
where $\Gamma$ is the decay width for
the  $\chi$-particles.
{}From Eq. (\ref{kinetic}) the second term on the
right is proportional to ${\dot \varphi}$, which in Eq. (\ref{eqnew})
leads to dissipative effects on $\varphi$ from its interaction with the
$\chi$-fields.  Analogous expressions also apply to the
fermionic averages (for Fermi-Dirac statistics) in Eq. (\ref{eqnew}).

Based on a systematic perturbative approach,
as we have shown in previous work \cite{GR,BGR},
we can write Eq. (\ref{eqnew}), using 
the expressions for the associated averages for the $\chi_{ij}$,
$\psi$ and $\eta$ fields
and  Eq. (\ref{kinetic}), as

\begin{equation}
\ddot{\varphi} + 3 H \dot{\varphi} + V^\prime_{\rm eff} (\varphi,T)
+ \eta (\varphi) \dot{\varphi} = 0 \;,
\label{eqVeff}
\end{equation}

\noindent
where $V^\prime_{\rm eff} (\varphi,T)= \partial V_{\rm
eff} (\varphi,T)/\partial \varphi$, is the field derivative of the
1-loop finite temperature
effective potential, which can be computed by the standard methods
(in fact this is the resummed effective potential, with masses written
in terms of the finite temperature effective masses) and
$\eta (\varphi) \equiv \eta^{\rm B}(\varphi)+\eta^{\rm F}(\varphi)$
is a field dependent dissipation, 

\begin{eqnarray}
\lefteqn{\eta^{\rm B} (\varphi) = \frac{\lambda^2}{8} \varphi^2 
\frac{1}{\beta^{-1}}
\int \frac{d^3 q}{(2 \pi)^3}
\frac{n_\phi^{\rm eq.}
(1 + n_\phi^{\rm eq.})}{\omega_\phi^2 ({\bf q}) \Gamma_\phi ({\bf q})} }
\nonumber \\
& & + \sum_{i=1}^{N_M} \sum_{j=1}^{N_{\chi}} \frac{g^4}{2} 
\frac{(\varphi-M_i)^2}
{\beta^{-1}} \int \frac{d^3 q}{(2 \pi)^3}
\frac{n_{\chi_{ij}}^{\rm eq.} (1 + n_{\chi_{ij}}^{\rm eq.})}
{\omega_{\chi_{ij}}^2 ({\bf q})
\Gamma_{\chi_{ij}} ({\bf q})}  
\label{dissbcoef}
\end{eqnarray}
and
\begin{equation}
\! \! \eta^{\rm F}(\varphi)= \!
\sum_{i=1}^{N_M} \sum_{j=1}^{\frac{N_{\chi}}{4}} g^2 \frac{(\varphi-M_i)^2}{
\beta^{-1}} \! \! \int \! \! \frac{d^3 q}{(2 \pi)^3}
\frac{n_{\psi_{ij}}^{\rm eq.} (1 - n_{\psi_{ij}}^{\rm eq.})}
{\omega_{\psi_{ij}}^2 ({\bf q})
\Gamma_{\psi_{ij}} ({\bf q})}.
\label{dissfcoef}
\end{equation}

\noindent
In what follows, we will be interested in the regime where
$\eta(\varphi) \gg 3 H$ in Eq. (\ref{eqVeff}). As discussed in 
\cite{hosoya,BGR,YL}, Eq. (\ref{eqVeff}) is valid provided the
adiabatic regime for $\varphi$ is satisfied, {\it i.e.},
the dynamic time-scale for $\varphi$ must be
much larger than the typical collision time-scale ($\sim \Gamma^{-1}$),
or $\left| \varphi/\dot{\varphi} \right| \gg \Gamma^{-1}$,
where $\Gamma$ is the smallest of the thermal
decay widths $\Gamma_\phi, \; \Gamma_{\chi_{ij}}, 
\; \Gamma_{\psi_{ij}}$, as it will set the
largest time-scale for collisions for the system in interaction with
the bath. 

Note that the damping coefficient depends on $\varphi^2$
or $(\varphi-M_i)^2$. To obtain
enough inflation, the amplitude of the inflaton should decrease very 
slowly, which requires efficient overdamping. This
overdamping is guaranteed by having a succession of fields always
thermalized, so that the population of decay products is not
depleted by the expansion. This condition is satisfied through our choice
of Lagrangian for the DM-model, which naturally guarantees that a population
of decay products will be generating efficient damping as $\varphi$ slowly
rolls down. 

We construct a warm inflation scenario based on the following
DM-model. For definiteness, we will refer all dimensional
quantities to $T_{BI}$, the temperature of the universe at the
beginning of warm inflation. The crucial property of the DM model of Eq.
(\ref{Nfields}) is that for a given temperature $T$, only the fields 
with masses $g^2(\varphi - M_i)^2 \stackrel{<}{\sim} T^2$ will contribute
to the thermal viscosity; the effect of heavier fields can be 
neglected.
Thus, as the inflaton rolls down its potential, we only must consider
the subset of fields for values of $i$ which satisfy the above inequality. 
Note that with this choice of model,
if $\varphi \gg T_{BI}$, which
is needed for efficient inflation,
it is still possible to have many light $\chi_{ij}$ $(\psi_{ij})$ fields if
$\varphi \sim M_i$. This will guarantee an efficient viscosity term in the
equation for $\varphi$.

{}For convenience, we write the mass parameters as $M_i = (i+i_{\rm min})
{\bar m}_{\chi \psi}^{\rm max} T_{BI}/g$, 
$i=1,\ldots,N_M$, with ${\bar m}_{\chi \psi}^{\rm max}$ a dimensionless
constant of ${\cal O} (1)$ and 
$M_{i=0} = i_{\rm min} T_{BI}/g$ the lowest mass level
coupled to $\varphi$.
{}For such a model, at $T \sim
T_{BI}$ and $M_{i+1} > \varphi > M_i$, there will be a range of masses 
when $2.5N_{\chi}+1$
$\chi_{ij},\psi_{ij},\chi_{i+1,j},\psi_{i+1,j}$, 
and $\eta$ fields are thermally excited. All
other $\chi_{i'j} (\psi_{i'j})$-fields, for $i' \ne i,i+1$ and 
$j=1, \ldots N_{\chi}\; (N_{\chi}/4)$, have
masses $m_{\chi \psi}^2 \approx g^2(\varphi-M_{i'})^2 > 
({\bar m}_{\chi \psi}^{\rm max} T_{BI})^2 \sim T^2$ 
and are thus cold (actually thermal excitation begins once
$m_{\chi \psi} \approx 10T$ and nontrivial dissipation 
once $m_{\chi \psi} \approx (2-3)T$). 

{}For the $\eta-\chi-\psi$ system participating in the dissipative
dynamics of $\varphi$, in each interval $M_{i+1} > \varphi > M_i$,
the radiation energy density is 
$(2N_{\chi}+7N_{\psi}+ \! 1) \pi^2 T^4/30$. 
In addition to these fields,
in general there can be a set of 
``heat bath'' fields that increase
the particle degrees in the radiation system, but do not otherwise
contribute to dissipation. {}For later use, we addopt
the following notation for the number of these heat bath fields,
$2(15N_{\chi}/8+ \! 1)^{1+{\alpha}} \! - (15N_{\chi}/4+ \! 1)$, so that,
in total, the radiation system has 
$g_* \equiv 2(15N_{\chi}/8+ \! 1)^{1+\alpha}$
particle degrees for any interval where $\varphi$ is. 
$\alpha \geq 0$
is a free parameter.

To simplify the
calculation to follow, we consider the region where $\varphi \gg T$
and the $\lambda \varphi^3/6$ term dominates the equation of motion.
{}From $V^\prime_{\rm eff} (\varphi,T)$ the leading $\chi,\psi$ field
contribution for $M_{i+1}> \varphi > M_i$ is
$(N_{\chi}T_{BI}^2/8) g^2 [(\varphi-M_i)+(\varphi-M_{i+1})]<
{\cal O} (g N_{\chi} T^3)$, so that the constraint requires $\lambda
\varphi^3 \gg {\cal O}(g N_{\chi} T^3)$. 
In the perturbative regime that we examine, $\lambda
\ll 1$, so that 
$m_{\phi}^2(\varphi,T) \approx \lambda \varphi^2/2 \equiv {\bar
m}_{\phi}^2(\phi,T) T_{BI}^2$,
since $\varphi \gg T$. The $\chi (\psi)$-masses will 
range from $f^2T^2/12$ $(g^2T^2/6) <
m_{\chi}^2 (m_{\psi}^2) < f^2T^2/12$ $(g^2T^2/6) + g^2(\varphi-M_i)^2 
\stackrel{<}{\sim} {\cal O}(T_{BI}^2)$. As a
simplification, the $\chi,\psi$-masses will be estimated at 
their largest value
$(m_{\chi \psi}^{\rm max})^2 \equiv 
({\bar m}_{\chi \psi}^{\rm max} T_{BI})^2 \approx g^2(\varphi-M_i)^2 
|_{\rm max}\approx {\cal O}(T^2_{BI})$. 
We can then express the condition that 
the $\lambda \varphi^3/6$ term dominates the equation of
motion Eq. (\ref{eqVeff}), in terms of 

\begin{equation}
R_{\chi \psi/ \varphi} \leq 
\frac{3g N_{\chi} m_{\chi \psi}^{\rm max}
T_{BI}^2}{4\lambda \varphi_{BI}^3} < 1\;,
\label{rchiphi2}
\end{equation}

\noindent
where, $R_{\chi \psi/ \varphi} \equiv
{{3g^2 N_{\chi} T^2_{BI} \left[(\varphi-M_i)+(\varphi-M_{i+1})
\right]_{\rm max}}\over {
4\lambda \varphi_{BI}^3}}$.

\noindent
To impose the most stringent constraint from this, it
will be taken at the maximum value $m_{\chi\psi}^{\rm max}$ from Eq.
(\ref{rchiphi2}).
In fact, considerable increase in dissipation, thus
improvement in the results to follow, occur by accounting for
corrections when the $\chi$'s ($\psi$'s) 
are in the smaller mass region. 
In the regime outlined above, the effective
equation of motion for $\varphi$, Eq. (\ref{eqVeff}), in the interval
$M_{i+1} > \varphi > M_{i}$ and in the overdamped
limit is

\begin{equation}
\eta_{i,i+1}(\varphi) {\dot \varphi} \simeq
-\frac{\lambda}{6} \varphi^3\;,
\label{phieom}
\end{equation}

\noindent
where $\eta_{i,i+1}(\varphi) \equiv 
\eta^{\rm B}_1 \left[ (\varphi-M_i)^2 + (\varphi-M_{i+1})^2 \right]
+\eta_1^{\rm F} T^2$
where 
$\eta^{\rm B}_1$ and $\eta^{\rm F}_1$ (from \cite{BGR} and \cite{YL})
are given by: $\eta_1^B \sim 384 N_\chi g^4 /[\pi T (f^2 + 8 g^4)]
\ln (2 T / m_\chi)$ and
$\eta_1^F \sim 11 N_\psi/T$, respectively.

As $\varphi$ moves through the interval
$M_{i+1} > \varphi > M_{i}$,

\begin{equation}
\eta_{i,i+1}(\varphi) = 
\kappa \left(1+{\rm r}_{\rm FB}\right) \eta^{\rm B}_1 ({\bar m}_{\chi \psi}^{\rm max} T_{BI})^2/ g^2
\;,
\label{bareta}
\end{equation}

\noindent
with
${\rm r}_{\rm FB} \equiv \eta^{\rm F}_1T_{BI}^2 g^2/[\eta^{\rm B}_1 
\kappa({\bar
m}_{\chi \psi}^{\rm max}T_{BI})^2] \approx 0.2$,
where the estimate on the right is from \cite{BGR,YL} for
the high temperature limit and  $0.5 < \kappa < 1$. 
Since we are examining the region $\varphi \gg
T$, we can ignore the weak $\varphi$ dependence in $\eta_{i,i+1}$.
In this case, the solution to Eq. (\ref{phieom}) is

\begin{equation}
\varphi(\tau) = \varphi_{BI} \left[ \frac{\tau+\tau_0}{W_4} +1 \right]^{-1/2},
\label{varphi}
\end{equation}

\noindent
with $W_4=3H_{BI} \kappa \eta^{\rm B}_1 
({\bar m}_{\chi \psi}^{\rm max} T_{BI})^2(1+{\rm r}_{\rm FB})
/(g^2 \lambda \varphi_{BI}^2)$ and
$H_{BI} = \sqrt{2\pi \lambda \phi_{BI}^4 /(9m_p^2)}$.
Time has been rescaled as $t=\tau/H_{BI}$ with
the origin chosen
such that $\varphi(\tau=0) \equiv \varphi_{BI}$, which implies
$\tau_0=0$.

The resulting warm inflation cosmology for such a field trajectory in
any interval $M_{i+1} > \varphi > M_{i}$ is equivalent to the $n=4$
model in \cite{ab2}. It yields a power-law warm inflation
which never terminates and for which in the steady state regime
$\rho_r(t)/\rho_v(t) = {\rm const}$. In our model, warm inflation
terminates into a radiation dominated regime 
once $\varphi < M_{i=0}$,
since below that point the dissipative 
term becomes negligible, in which
case $\varphi$ coasts down the potential. 
The essential point of interest here is to show that
once $\varphi$ reaches $M_{0}$, sufficient ${\rm e}$-folds $N_e$ have occurred
while the universe has nonnegligible radiation.

{}For simplicity, the steady state cosmology 
in flat spatial geometry will be examined, which
implies from \cite{ab2}, for $W_4 \gg 1$, 
$\rho_r(0)/\rho_v(0)
=\rho_r(\tau)/\rho_v(\tau) = 1/(2W_4)$.
[{}From Eq. (\ref{varphi}),
we can show
that this is the necessary condition to satisfy 
the adiabatic condition, $\left| \varphi/\dot{\varphi} \right| \gg
\Gamma^{-1} $ and $\Gamma_{\phi(\chi)} > H$.]
In terms of  the parameters of the model this can also be written as
$g_* \pi^2 T_{BI}^4 /30 = 
\lambda  \varphi^4_{BI}/(48 W_4)$.
Initial conditions that are more realistic, such as $\rho_r(0) >
\rho_v(0)$ rapidly evolve into the steady state behavior. In this steady
state regime, the scale factor solution is \cite{ab2}

\begin{equation}
R(\tau) = \left(\tau/W_4 + 1 \right)^{W_4+\frac{1}{4} }\;,
\label{plscale}
\end{equation}

\noindent
with initialization $R(0)=1$.

Our goal is to compute $W_4$ from the microscopic parameters of the
model and consistent with the many constraints given above and in
section V of \cite{BGR}. The power-law expansion behavior of the scale
factor, Eq. (\ref{plscale}), is such that the major factor of growth
happens for $\tau/W_4 < 10$. $N_{\rm e}=W_4$ ${\rm e}$-folds 
occur at time $\tau / W_4
= {\rm e}-1$. We will restrict our 
calculation within this time interval. At
the end of this interval $\varphi(\tau)$, Eq. (\ref{varphi}), and $T$
have not changed significantly, $\varphi_{BI} \rightarrow
\varphi_{BI}/\sqrt{{\rm e}}$,
$T_{BI} \rightarrow T_{BI}/\sqrt{{\rm e}}$. This
simplifies the constraint equations, since they can be
analyzed at the initial values $\varphi_{BI}$, $T_{BI}$ and
approximately will be valid throughout this interval.

The constraint equations for computing $W_4$ are as follows.
To satisfy
the thermalization conditions 
$\Gamma_{\chi},\Gamma_{\phi}, \Gamma_{\psi} > H$, we will
set $H_{BI}={\rm min}(\Gamma_{\chi},\Gamma_{\phi},\Gamma_{\psi})$. 
More explicitly, for the warm inflation solutions studied below,
$\Gamma_{\chi}$ (which is the smallest of the $\Gamma$s)
is larger than $H_{BI}$ for $f \stackrel{>}{\sim} 0.8$.
This condition may
be under restrictive to obtain good thermalization, but it provides the
maximum parameter region that may be useful. 
It should be noted that since $H \sim \varphi^2$ and
$\Gamma \sim T$, as warm inflation proceeds, thermalization improves. 
The thermalization condition automatically implies that the adiabatic
condition is comfortably satisfied.

All the constraints are expressed in the following four relations

\begin{equation}
R_{\chi\psi/\varphi}^2 
= 
\frac{15 g^4 N_{\chi}^2 \left(\frac{15N_{\chi}}{8}+1\right)^{-(1+\alpha)}  }
{256 \pi^2 \kappa [{\rm min}(\Gamma_\phi,\Gamma_{\chi})]
\eta^{\rm B}_1(1+{\rm r}_{\rm FB}) }\;,
\label{rnalprcp2}
\end{equation}

\begin{equation}
W_4 = \frac{45 \left(g N_{\chi} {\bar m}_{\chi \psi}^{\rm max}\right)^2}
{512 \pi^2 {\bar m}_{\phi_{BI}}^2 R_{\chi \psi/\varphi}^2 
\left(\frac{15N_{\chi}}{8}+1\right)^{1+\alpha}}\;,
\end{equation}

\begin{equation}
R_{\chi\psi/\varphi} \varphi_{BI}/T_{BI}=
3gN_{\chi}{\bar m}_{\chi \psi}^{\rm max}/
\left(8{\bar m}_{\phi_{BI}}^2\right)
\label{phiot}
\end{equation}

\noindent
and

\begin{equation}
\lambda R_{\chi\psi/\varphi}^{-2} =
128{\bar m}_{\phi_{BI}}^6
\left(3 gN_{\chi}{\bar m}_{\chi \psi}^{\rm max} \right)^{-2}\;.
\end{equation}

\noindent 
Eq. (\ref{rnalprcp2}) is obtained from the relation for $W_4$
below Eq. (\ref{varphi}), where $\eta (\varphi)$ in Eq. (\ref{bareta}),
has been expressed in terms of $R_{\chi\phi}$, 
Eq. (\ref{rchiphi2}). The
procedure is to input $g,f,{\bar m}_{\chi \psi}^{\rm max}, {\bar
m}_{\phi_{BI}},N_{\chi}$ on the right-hand-side of the above four
equations, then obtain the left-hand-side of Eqs. (\ref{rnalprcp2}) and
(\ref{phiot}) from which the remaining two equations follow, up to a
choice for $R_{\chi \psi/\varphi}<1$. Note that the only cases requiring
additional heat bath fields ($\alpha>0$) are in parametric regimes when
the right-hand-side of Eq. (\ref{rnalprcp2}) is greater than 1, since we
require $R_{\chi \psi/\varphi} < 1$.

In \cite{BGR} analytic expressions were obtained for $\Gamma({\bf q})$
and $\eta_1$ in the simplified limit $|{\bf q}|=0$, 
$m_{\chi}, m_{\psi} \sim {\cal
O}( m_{\phi})$ (which here we call level 1) 
as well as in terms of the exact 2-loop
expressions, which must be computed numerically (level 2). The results in
{}Fig. 1  present both levels of approximation.
We used $g=f=0.9$, $\kappa=0.5$, $N_{\chi}=12$,
$N_{\psi}=N_{\chi}/4=3$
and $R_{\chi\psi/\varphi} \stackrel{<}{\sim} 1$ 
($\alpha \stackrel{>}{\sim} 0$).
The solid and dashed curves are for ${\bar m}_{\chi\psi}^{\rm max}=0.9$,
for level 1 and level 2, respectively, and the dotted curve is level
2 for ${\bar m}_{\chi \psi}^{\rm max}=2.5$. 
{}For both curves drawn in {}Fig. 1, 
going from 
${\bar m}_{\phi_{BI}} =0.002$ to $0.05$, the initial
field displacement $\varphi_{BI}/T_{BI}$ ranges as
$10^6 - 10^3$, $N_M$ ranges as
$10^5 - 10^2$ and $\lambda$ 
ranges as $\sim 10^{-17}-10^{-9}$.
In the regime we are considering, $i_{\rm min} \sim \varphi_{BI}/(gT_{BI})$
and for the above results, $i_{\rm min}$ ranges from $10^2-10^6$.
To obtain $N_{\rm e} \sim 60$ ${\rm e}$-folds of warm inflation
for all three cases,
$\varphi_{BI}/T_{BI} \approx 3000$, $N_M=1000$, $\lambda \approx 10^{-9}$.
An absolute scale is fixed by setting $m_p=10^{19} {\rm GeV}$
from which for $N_{\rm e} \sim 60$ ${\rm e}$-folds, we find 
$T_{BI} \sim (10^{13}-10^{14}) {\rm GeV}$ and 
$H_{BI} \sim (10^9-10^{10}) {\rm GeV}$. [The temperature at the
onset of warm inflation, $\rho_v=\rho_r$, 
is $(2W_4)^{1/4} \sim 3.3$
times bigger than $T_{BI}$, and rapidly decreases to $T_{BI}$
during the transient period.]

The DM-model studied here was motivated by
the requirements of warm inflation, dissipative dynamics and
perturbative renormalizability. We can justify our choice of Lagrangian by
noting that string-inspired models can display an inflaton coupled to
mass modes of a string, as explained in Ref. \cite{ak}. Within this context,
the large number of fields necessary to realize sufficient inflation
is a natural consequence of the modifications of short-distance physics
required by string theories.

In summary, first principles quantum field interactions can realize a
warm inflation regime with sufficient duration to solve the
horizon/flatness problems. The interplay between inflationary expansion
and radiation production has been a persistent problem since the
earliest history of inflationary cosmology. Thus, despite the many
questions opened by our model, its underlying mechanism is a
unique and simple resolution to the problem. {}Further
study of the inflaton $k$-modes is necessary to address the
density fluctuation problem.

We thank A. Linde and J. Yokoyama for 
their interest.
AB was supported by the U.S. Department of Energy.
MG was partially supported by the NSF through a Presidential
Faculty Fellows Award no. PHY-9453431 and by a NASA
grant no. NAG5-6613. ROR was partially supported by CNPq and FAPERJ.

FIGURE CAPTIONS

{\bf Figure 1}: Number of ${\rm e}$-folds of warm 
inflation $N_{\rm e}$ versus
$m_{\phi_{BI}}$ for ${\bar m}_{\chi \psi}^{\rm max}=0.9$ level 1
(solid), $0.9$ level 2 (dashed), and $2.5$ level 2
(dotted) with $g=f=0.9$, $\kappa=0.5$, $N_{\chi}=12$,
and $N_{\psi}=N_{\chi}/4=3$.

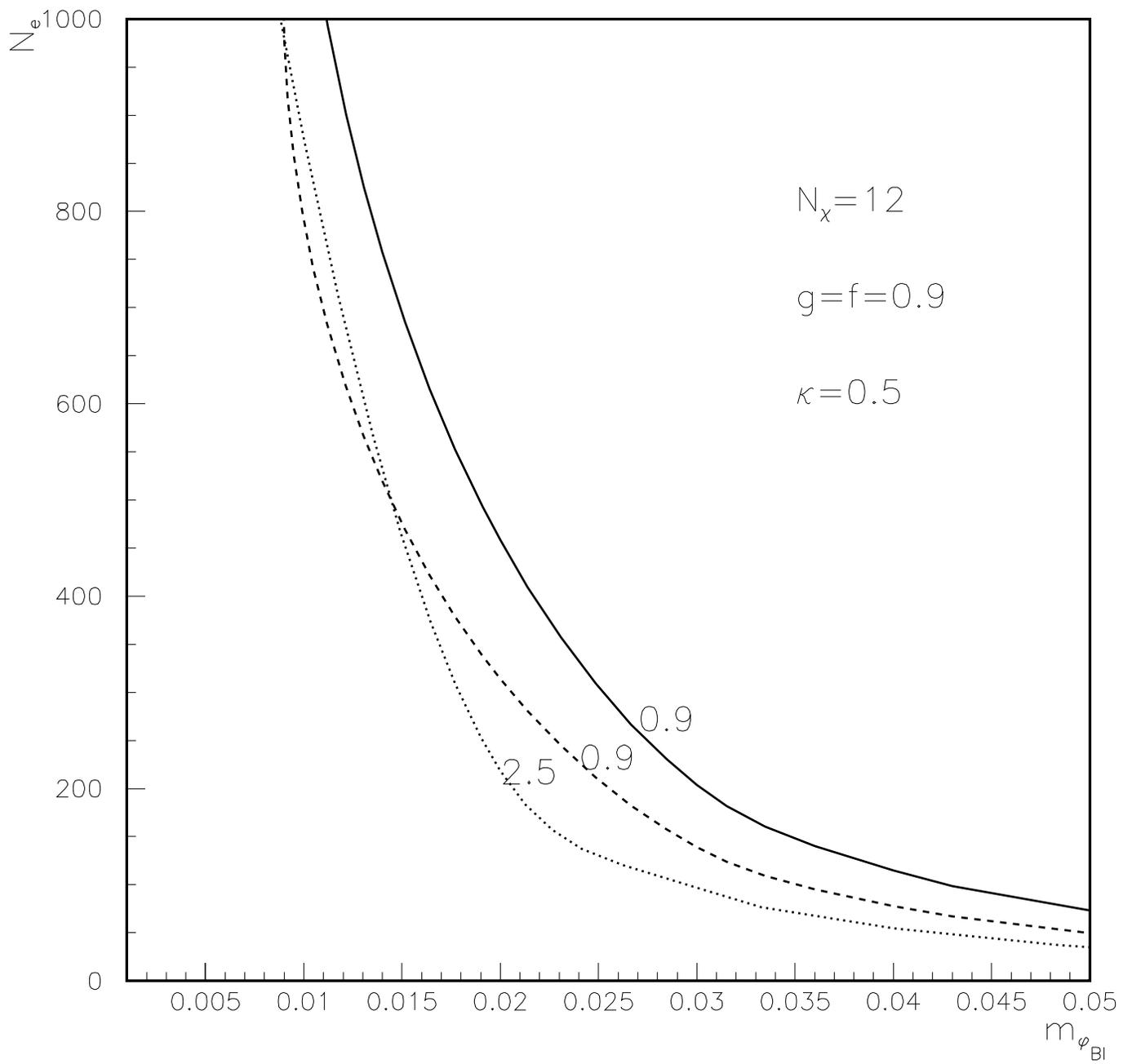
\end{document}